\documentclass[12pt,dvips]{article}
\usepackage[dvips]{graphicx}
\usepackage{graphicx}

\graphicspath{{plots/}}
\DeclareGraphicsExtensions{.eps.gz,.eps,.ps,.ps.gz}

\parskip=\itemsep               %?
\newcommand{\gwig}{\mbox{\,\raisebox{.3ex}
    {$>$}$\!\!\!\!\!$\raisebox{-.9ex}{$\sim$}}\,}
\newcommand{\lambdabar}{{\hbox{$\lambda_e$\kern-1.9ex\raise+0.45ex\hbox{--}
\kern+0.2ex}}}

\date{\empty}

\title{{\normalsize\rightline{DESY 01-182}\rightline{hep-ph/0111042}}
\vskip 1cm 
\bf Collider versus Cosmic Ray  Sensitivity to Black Hole Production
       \vspace{21mm}} 
\author{A. Ringwald and H. Tu\\[4mm] 
Deutsches Elektronen-Synchrotron DESY, Hamburg, Germany}

\begin{document}
\begin{titlepage} 
  \maketitle
% declarations for front matter
\vspace{1.5cm}
\begin{abstract}
In scenarios with 
extra dimensions and 
TeV-scale quantum gravity,
black holes are expected to be produced copiously at center-of-mass energies above
the fundamental Planck scale. 
The Large Hadron Collider (LHC) may thus turn into a factory of black holes, at which their
production and evaporation may be studied in detail. 
But even before the LHC starts operating, the Pierre Auger Observatory for cosmic rays, presently
under construction, has an opportunity to search for black hole signatures.  
Black hole production in the scattering of 
ultrahigh energy cosmic neutrinos on nucleons in the atmosphere may initiate 
quasi-horizontal air showers with distinct characteristics above the Standard Model rate. 
In this letter, we compare the sensitivity of LHC and Auger to black hole production by studying
their respective reach in black hole production parameter space. 
%rev
Moreover, we present constraints   
in this parameter space from the non-observation of horizontal showers by the Fly's Eye collaboration.
We find that if the ultrahigh energy neutrino flux is at the level expected from cosmic ray interactions with the
cosmic microwave background radiation, Auger has only a small window of opportunity to 
detect black holes before the start of the LHC. 
If, on the other hand, larger ultrahigh energy neutrino fluxes on the level of the upper limit from 
``hidden'' hadronic astrophysical sources are realized in nature, then the first signs 
of black hole production may be observed at Auger.
%rev
Moreover, in this case, the Fly's Eye constraints, although more model-dependent, turn out to be competitive with other 
currently available constraints on TeV-scale gravity which are mainly based on interactions associated with
Kaluza-Klein gravitons.
\end{abstract}


\thispagestyle{empty}
\end{titlepage}
\newpage \setcounter{page}{2}

%\section{Introduction}
{\em 1.}
It has been conjectured quite some time ago that black holes will be produced in the 
collision of two light particles at center-of-mass (cm) energies above the Planck scale
with small impact
%rev 
parameters~\cite{'tHooft:1987rb%,Amati:1987wq,
%'tHooft:1988wk,Amati:1988uf,Amati:1989tn,Amati:1990xe,Amati:1992zb,Amati:1993tb,
%Aref'eva:1995qs
}. 
This remote possibility seems now within reach  in the context of 
%Among the most spectacular consequences of 
theories with
$\delta = D-4\geq 1$ %large, compact 
flat~\cite{Arkani-Hamed:1998rs%
%,Antoniadis:1998ig,Arkani-Hamed:1999nn,
} or warped~\cite{Randall:1999ee} extra dimensions 
and a low fundamental Planck scale 
$M_D\,\gwig$ TeV characterizing quantum 
gravity.
In these theories one might expect the copious production of black holes 
in high energy collisions
at cm energies above 
$M_D$~\cite{Argyres:1998qn,%
%Banks:1999gd,Aref'eva:1999bm,
Emparan:2000rs,Giddings:2000ay%,Emparan:2001ce
}.
 
Recently it has been  
emphasized~\cite{Giddings:2001bu,Dimopoulos:2001hw} that 
the Large Hadron Collider (LHC)~\cite{Evans:2001mn}, 
expected  to have a first physics run in  2006, may turn into a factory of black holes
at which their production and evaporation may be studied in detail (see also 
Refs.~\cite{Dimopoulos:2001qe,Hossenfelder:2001dn%
%,Giddings:2001ih,Cheung:2001ue,Casadio:2001wh
}). 

Black hole production and subsequent decay in the scattering of ultrahigh energy cosmic neutrinos
on nucleons in the atmosphere may initiate quasi-horizontal air showers
far above the Standard Model rate. Recently it was argued~\cite{Feng:2001ib} that the search for
such air showers at the Pierre Auger Observatory~\cite{Zavrtanik:2000zi%
%,Zepeda:2000zk
} for extensive air showers, 
expected to be completed by the end of 2003, might have enough sensitivity to probe black hole physics if the 
fundamental Planck scale is below 2 TeV. The corresponding experimental signature was worked out in 
Ref.~\cite{Anchordoqui:2001ei}. Further discussions of cosmic ray issues associated with black 
hole production may be found in  Ref.~\cite{Emparan:2001kf}.
In  Ref.~\cite{Uehara:2001yk}, the detection of black holes at the planned neutrino telescope 
ICECUBE~\cite{Halzen:1999jy} was considered.

The purpose of the present letter is to compare the sensitivity of the LHC and Auger to black hole 
production by studying their respective reach in black hole production parameter space. 
Moreover, we derive constraints on this parameter space
from the non-observation of horizontal
showers~\cite{Baltrusaitis:1985mt} by the Fly's Eye collaboration~\cite{Baltrusaitis:1985mx}. 
These constraints on TeV-scale gravity complement the ones which arise from the confrontation of 
collider~\cite{Giudice:1999ck,Mirabelli:1999rt%
%,Han:1999sg,Hewett:1999sn,Mathews:1999kf,%
%Rizzo:1999fm,Mathews:1999qn,Cheung:1999qh,Dudas:2000gz,Cullen:2000ef,Accomando:2000sj,Adloff:2000dp
%,Davoudiasl:2000jd
}, astrophysical~\cite{Cullen:1999hc%
%,Barger:1999jf,Hanhart:2001er,Hanhart:2001fx,Hannestad:2001jv,Hannestad:2001xi
}, cosmological~\cite{Hall:1999mk%
%,Fairbairn:2001ct,Hannestad:2001nq
},
and cosmic ray~\cite{Nussinov:1999jt,%
%Jain:2000pu,
Tyler:2001gt,Kachelriess:2000cb%
%,Anchordoqui:2001mk,Anchordoqui:2001uh,Alvarez-Muniz:2001mk,Kachelriess:2001jq
} data with predictions mainly based on  
interactions associated with Kaluza-Klein 
gravitons (for recent reviews, see e.\,g. Ref.~\cite{Peskin:2000ti%
%,Abe:2001nq
}), according to which a fundamental Planck scale as low as $M_D = {\mathcal O}(1)$ TeV is still allowed for $\delta\geq 4$ flat 
or $\delta\geq 1$ warped extra dimensions.

In Sect.~{\em 2}, we review the phenomenological model for
black hole production and decay in scenarios with TeV-scale gravity. We determine 
the contribution of black hole production to the proton-proton and  
neutrino-nucleon cross section, respectively, for various values of the model parameters.
The reach of the LHC in the black hole parameter space follows immediately from these considerations. 
In Sect.~{\em 3}, we determine the rate of quasi-horizontal air showers initiated
by neutrino-nucleon scattering into black holes expected at Auger. 
In order to be able to make a fair comparison of the reach of Auger versus the LHC to 
black hole production, we exploit both conservative lower and upper limits on the 
presently unknown flux of ultrahigh energy cosmic neutrinos.  
We recall also the Fly's Eye upper limit on the
ultrahigh energy neutrino flux times the neutrino-nucleon cross section and present various
upper limits on the neutrino-nucleon cross section, obtained from various 
assumptions about the neutrino flux. These upper limits are then 
turned into exclusion regions in the black hole parameter space.    
Section~{\em 4} contains our conclusions. 

%\section{Black hole phenomenology}
{\em 2.}
We start with a review of the current understanding of the production and decay 
of black holes in TeV-scale gravity 
scenarios~\cite{Argyres:1998qn,%
%Banks:1999gd,Aref'eva:1999bm,
Emparan:2000rs,Giddings:2000ay,%Emparan:2001ce,
Giddings:2001bu,Dimopoulos:2001hw,%
Dimopoulos:2001qe,Hossenfelder:2001dn%
%,Giddings:2001ih,Cheung:2001ue
}.

Based on semiclassical
%rev 
reasoning~\cite{'tHooft:1987rb%,Amati:1987wq,%
%'tHooft:1988wk,Amati:1988uf,Amati:1989tn,Amati:1990xe,Amati:1992zb,Amati:1993tb,
%Aref'eva:1995qs
}, one expects that at trans-Planckian parton-parton cm energies squared, $\hat s\gg M_D^2$,  
and at small parton-parton impact parameters, $b\ll r_S\,(M_{\rm bh}=\sqrt{\hat s})$, i.\,e. 
at impact parameters much smaller than the Schwarzschild radius $r_S$ of a $(4+\delta )$-dimensional black 
hole with mass $M_{\rm bh}=\sqrt{\hat s}$~\cite{Myers:1986un}\footnote{We define $M_D$ as in 
Ref.~\cite{Giudice:1999ck}. Equation~(\ref{schwarzsch}) is valid as long as $r_S\ll R_c$, with $R_c$ being the compactification 
or curvature radii in the flat or warped scenario, respectively.}, 
\begin{equation}
\label{schwarzsch}
r_S =\frac{1}{M_D}
\left[
\frac{M_{\rm bh}}{M_D}
\left(
\frac{2^\delta \pi^{\frac{\delta -3}{2}}\,\Gamma\left( \frac{3+\delta}{2}\right)}{2+\delta}
\right)
\right]^{\frac{1}{1+\delta}}
\,,
\end{equation} 
a black hole forms with a cross section%\footnote{}
\begin{equation}
\label{sig_bh_geom}
\hat\sigma (ij\to {\rm bh})\equiv  
\hat\sigma^{\rm bh}_{ij} (\hat s ) \approx \pi\,r_S^2 
\left( M_{\rm bh}=\sqrt{\hat s} \right)\,
\theta\left( \sqrt{\hat s} -M_{\rm bh}^{\rm min}\right) 
\,.
\end{equation}
Here, $M_{\rm bh}^{\rm min}\gg M_D$ parametrizes the cm energy above which the
semiclassical reasoning mentioned above is assumed to be valid.

Some caveats have to be mentioned, however.   
As noted in Ref.~\cite{Voloshin:2001vs}, the semiclassical production of black
holes resembles largely the problem of baryon and lepton number violating processes 
(``sphaleron~\cite{Klinkhamer:1984di} production'') in multi-TeV ($\sqrt{\hat s}\gg m_W/\alpha_W$) 
particle collisions in the standard electroweak theory~\cite{Ringwald:1990ee,%
%Espinosa:1990qn,Farrar:1990vb,
Ringwald:1991bh,%
Ringwald:1991qz,
Morris:1991bb,Morris:1994wg,Gibbs:1995cw},
which is not yet completely understood~\cite{Mattis:1992bj%
%,Tinyakov:1993dr,Voloshin:1994yp,Rubakov:1996vz
}.  
In fact, also in the latter case a simple geometric behavior similar to Eq.~(\ref{sig_bh_geom}), with
the Schwarzschild radius replaced by the sphaleron radius $\sim m_W^{-1}$,  
was advocated in Refs.~\cite{Ringwald:1991bh,Ringwald:1991qz}.
In both cases, there might be additional exponential suppression factors rendering semiclassical 
sphaleron or black hole production unobservable in the TeV range~\cite{ Voloshin:2001vs} 
(see, however, Ref.~\cite{Dimopoulos:2001qe}).

With these caveats understood, one may infer from the estimate~(\ref{sig_bh_geom}) the contribution of 
black hole production to the proton-proton and neutrino-nucleon  cross section, 
\begin{eqnarray}
\label{sig_pp_bh_pdf}
\sigma_{pp}^{\rm bh}(s) &=&
\sum_{ij} \int^1_0 {\rm d}x_1\,{\rm d}x_2\,
\frac{f_i (x_1,\mu )\,f_j (x_2,\mu )+f_i (x_2,\mu )\,f_j (x_1,\mu )}
{1+\delta_{ij}}\,\hat \sigma_{ij}^{\rm bh} (x_1\,x_2\,s)
\,, 
\\
\label{sig_nuN_bh_pdf}
\sigma_{\nu N}^{\rm bh} (s) &=&
\sum_i \int^1_0%_{M_{\rm bh}^{{\rm min}\,2}/s}^1
{\rm d}x\,f_i (x,\mu )\,\hat \sigma_{\nu i}^{\rm bh} (xs)
\, .
\end{eqnarray}
Here, $s$ denotes the proton-proton  or  neutrino-nucleon cm  energy squared. The sum 
extends over all partons in the nucleon, with parton distribution functions $f_i(x,\mu )$ and 
factorization scale $\mu$.
For our numerical integration we have used various sets of parton distributions as they are 
implemented in the parton distribution library 
PDFLIB~\cite{Plothow-Besch:1993qj%
%,Plothow-Besch:1995ci,Plothow-Besch:2001
}. Uncertainties associated with different parton distribution sets 
are in the ${\mathcal O}(20)\, \%$ range and are not explicitely displayed in the following.

%%%%%%%%%%%%%%%%%%%%%%%%%%%%%%%%FIGURE%%%%%%%%%%%%%%%%%%%%%%%%%%
\begin{figure}
\begin{center}
\parbox{12cm}{\includegraphics*[bbllx=20pt,bblly=221pt,bburx=570pt,bbury=608pt,width=12cm]{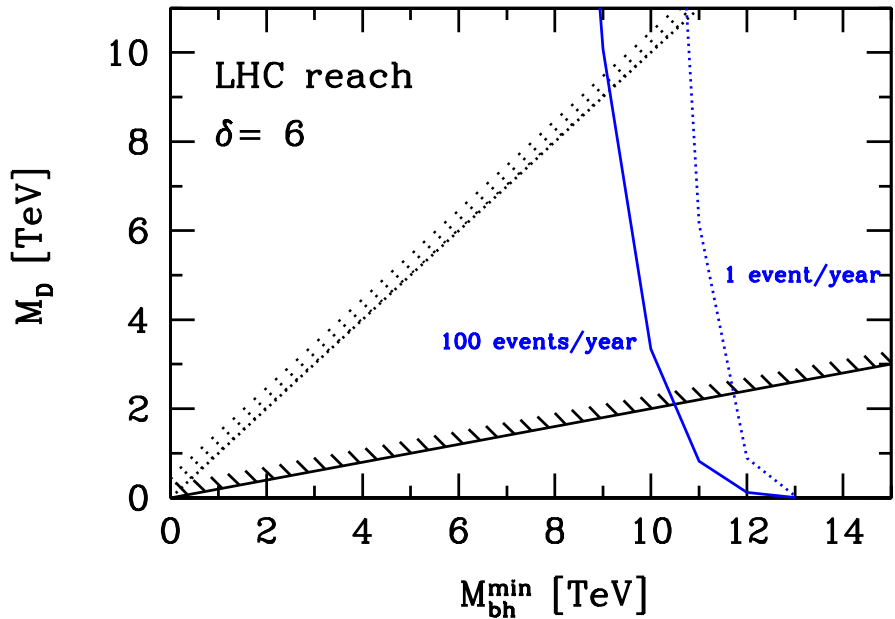}}
\caption[dum]{\label{bh_par_lhc_d6}
Accessible region in the black hole production parameters at the LHC for $\delta =6$ extra dimensions.  
The solid and the dotted lines are 
contours of constant numbers of produced black holes per year ($10^7$ s) with a mass larger than $M_{\rm bh}^{\rm min}$, 
for a fundamental Planck mass $M_D$. The shaded dotted, $M_D = M_{\rm bh}^{\rm min}$, and shaded solid, 
$M_D=(1/5)\,M_{\rm bh}^{\rm min}$, lines give a rough indication of the boundary of applicability of  the semiclassical 
picture~\cite{Giddings:2001bu}.  
}
\end{center}
\end{figure}
%%%%%%%%%%%%%%%%%%%%%%%%%%%%%%%%%%%%%%%%%%%%%%%%%%%%%%%%%%%%%%%%%

The reach of the LHC to black hole production is illustrated in Fig.~\ref{bh_par_lhc_d6} for $\delta = 6$ extra 
dimensions.   
%-- the reach for smaller $\delta$ being somewhat larger.  
The number of black hole events produced in a time interval $\triangle t$, 
$N_{\rm bh} =  \sigma_{pp}^{\rm bh} \cdot {\mathcal L}\cdot  \triangle t$,  has been calculated using the 
CTEQ5D parton distributions~\cite{Lai:2000wy} with\footnote{\label{foot_fact} If one
uses, instead, the other natural factorization scale $\mu = r_S^{-1}$~\cite{Giddings:2001bu,Emparan:2001kf}, which is 
typically much smaller
than $\sqrt{\hat s}$, the predicted production rates decrease by a factor of ${\mathcal O}(2)$.}
 $\mu = {\rm min}\,(\sqrt{\hat s},10\ {\rm TeV})$ in 
Eq.~(\ref{sig_pp_bh_pdf}) and the LHC design values $\sqrt{s}=14$ TeV
for the proton-proton cm energy and ${\mathcal L}=10^{34}$ cm$^{-2}$\,s$^{-1}$ for the luminosity~\cite{Evans:2001mn}.
As can be seen from Fig.~\ref{bh_par_lhc_d6}, the LHC can explore the production of black holes
with minimum masses nearly up to its kinematical limit of $14$ TeV, if $M_D$
is below $2$ TeV. 

%%%%%%%%%%%%%%%%%%%%%%%%%%%%%%%%FIGURE%%%%%%%%%%%%%%%%%%%%%%%%%%
\begin{figure}
\begin{center}
\parbox{12cm}{\includegraphics*[bbllx=20pt,bblly=221pt,bburx=570pt,bbury=608pt,width=12cm]{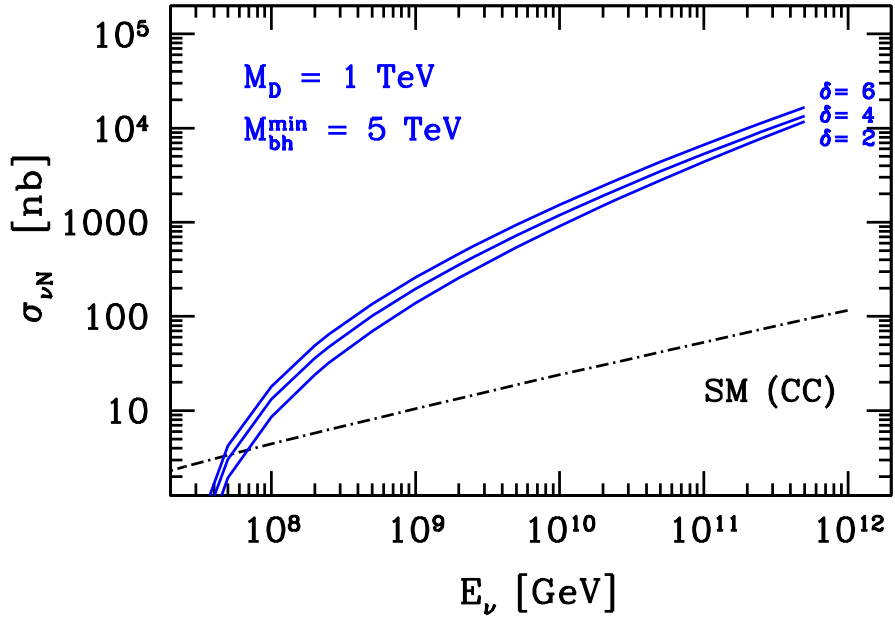}}
\caption[dum]{\label{sig_nuN_bh}
Cross section $\sigma_{\nu N}^{\rm bh}$, Eq.~(\ref{sig_nuN_bh_pdf}), for black hole production in 
neutrino-nucleon scattering, 
for $M_D = 1$ TeV, $M_{\rm bh}^{\rm min} = 5$ TeV, and $\delta = 2,4,6$ extra dimensions (solid lines, from bottom to top).   
Also shown is the Standard Model (SM) charged current (CC) neutrino-nucleon cross section (dashed-dotted line). 
}
\end{center}
\end{figure}
%%%%%%%%%%%%%%%%%%%%%%%%%%%%%%%%%%%%%%%%%%%%%%%%%%%%%%%%%%%%%%%%%

In order to appreciate the event numbers indicated in Fig.~\ref{bh_par_lhc_d6}, let us mention the 
expected signature of black hole decay, which is quite spectacular. Once produced, black holes decay primarily 
via Hawking radiation~\cite{Hawking:1975sw} into a large number of ${\mathcal O}(20)$ hard quanta, with energies 
approaching several hundreds of GeV. A substantial fraction of the beam energy is deposited in visible
transverse energy, in an event with high sphericity. From previous studies of sphaleron production, which has quite 
similar event characteristics~\cite{Ringwald:1991qz,Morris:1991bb,Morris:1994wg,Gibbs:1995cw}, as well as from first 
event simulations 
of black hole production~\cite{Dimopoulos:2001hw}, it is clear that only a handful of 
such events is needed at the LHC to discriminate them from perturbative Standard Model background.  

The contribution of black hole production to the neutrino-nucleon cross section is displayed in Fig.~\ref{sig_nuN_bh}, 
for $M_D = 1$ TeV, $M_{\rm bh}^{\rm min}=5$ TeV, and various values of $\delta$, as a function of 
the neutrino's energy in the nucleon's rest frame, $E_\nu = s/(2\,m_N)$, with $m_N=(m_p+m_n)/2$ being the nucleon mass.
Here, the CTEQ3D~\cite{Lai:1995bb} parton distributions with$^{{\ref{foot_fact}}}$ 
$\mu = {\rm min}\,(\sqrt{\hat s},10\ {\rm TeV})$ have been used in Eq.~(\ref{sig_pp_bh_pdf}).
The Standard Model charged current contribution, also shown in Fig.~\ref{sig_nuN_bh}, has been taken 
from Ref.~\cite{Gandhi:1998ri}, which compares
favorably with the one presented in Ref.~\cite{Gluck:1999js%
%,Kwiecinski:1999yf
}.
For the time being, we ignore possible unitarity corrections which might reduce the 
Standard Model contribution somewhat~\cite{Dicus:2001kb%
%,Kusenko:2001gj,Reno:2001hv
}.

%\section{Event rates at AUGER}
{\em 3.}
Let us consider now the rate of quasi-horizontal air showers initiated
by neutrino-nucleon scattering into black holes expected at the Pierre Auger Observatory.
For neutrino-nucleon cross sections below ${\mathcal O}(10)\ \mu$b, the neutrino flux attenuation in 
the upper atmosphere can be neglected, and the number of black hole initiated horizontal
air showers with an energy larger than a threshold energy $E_{\rm th}$ expected to be measured at Auger 
in a time interval $\triangle t$ is given by
\begin{equation}
\label{rate_auger}
N_{\rm sh}^{\rm bh}\, ( > E_{\rm th} ) = \triangle t\,N_A\,\rho_{\rm air}
\int_{E_{\rm th}}^\infty {\rm d}E_\nu\,F_\nu (E_\nu )\,\sigma_{\nu N}^{\rm bh}(E_\nu )\,
{\mathcal A}(E_\nu )\,,
\end{equation}
where $N_A$ is Avogadro's constant, $\rho_{\rm air}\simeq 10^{-3}$ g\,cm$^{-3}$ is the air density,
$F_\nu = \sum_i (F_{\nu_i}+F_{\bar\nu_i})$ is the sum of the differential diffuse neutrino fluxes, 
and $\mathcal A$ is the detector acceptance~\cite{Capelle:1998zz}. Note, that Eq.~(\ref{rate_auger}) assumes that 
100\,\% of the 
incident neutrino energy goes into 
visible, hadronic or electromagnetic shower energy, as it is the case for Standard Model, $\nu_e$ and 
$\bar\nu_e$ initiated charged current interactions, as well as  for 
sphaleron~\cite{Morris:1991bb,Morris:1994wg,Gibbs:1995cw} and 
black hole~\cite{Emparan:2000rs,Feng:2001ib} production and decay, at 
least to a good approximation.

\begin{figure}[ht]
\begin{center}
\includegraphics*[bbllx=20pt,bblly=221pt,bburx=570pt,bbury=608pt,%
width=12cm]{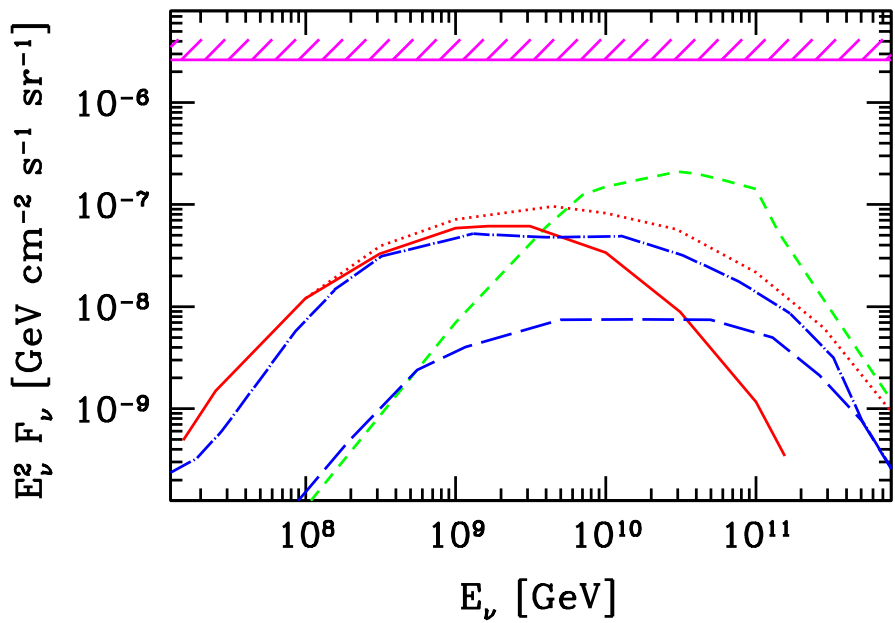}
\caption[dum]{\label{e2flux_cosmogenic}
Predictions of the cosmogenic neutrino flux, $F_\nu =\sum_i\left[ F_{\nu_i}+
F_{\bar\nu_i}\right]$. 
Short-dashed line: Flux from Ref.~\cite{Stecker:1979ah} (cf. Ref.~\cite{Feng:2001ib}). 
Long-dashed (long-dashed--dotted) line: Flux from 
Ref.~\cite{Yoshida:1993pt} for cosmological evolution parameters 
$m=2$, $z_{\rm max}=2$ ($m=4$, $z_{\rm max}=4$).
Solid (dotted) line: Flux from Ref.~\cite{Protheroe:1996ft}, 
assuming a maximum energy of $E_{\rm max}=3\cdot 10^{20(21)}$ eV for the ultrahigh energy cosmic rays.
Shaded: Theoretical upper limit of the neutrino flux from ``hidden'' %hadronic 
astrophysical sources that are
non-transparent to ultrahigh energy nucleons~\cite{Mannheim:2001wp}.
}
\end{center}
\end{figure}

Of central importance in the evaluation of the event rate~(\ref{rate_auger}) is
the expected differential flux $F_\nu$ of ultrahigh energy neutrinos to which we turn our 
attention next (for recent reviews, see  
Ref.~\cite{Protheroe:1999ei%
%,Gandhi:2000kq,Learned:2000sw
}).
Though atmospheric neutrinos, i.\,e. neutrinos produced in hadronic showers in the atmosphere, 
are certainly present, their flux in the ultrahigh energy region is anticipated
to be negligible~\cite{Volkova:1980sw%
%,Lipari:1993hd
}. 
Much more promising, but also more or less guaranteed are the so-called 
cosmogenic neutrinos which are produced when ultrahigh energy protons
inelastically scatter off the cosmic microwave background radiation~\cite{Greisen:1966jv%
%,Zatsepin:1966jv
}
in processes such as $p\gamma\to \Delta\to n\pi^+$, where the produced pion subsequently 
decays~\cite{Beresinsky:1969qj,%
%Beresinsky:1970,
Stecker:1979ah,Hill:1985mk%
%,Hill:1986fm,Stecker:1991vm
}.
Recent estimates of these fluxes  can be found in 
Refs.~\cite{Yoshida:1993pt,Protheroe:1996ft,Yoshida:1997ie,Engel:2001hd}, some of which 
are shown in Fig.~\ref{e2flux_cosmogenic}. 

%%%%%%%%%%%%%%%%%%%%%%%%%%%%%%%%FIGURE%%%%%%%%%%%%%%%%%%%%%%%%%%
%%%%%%%%%%%%%%%%%%%%%%%%%%%%%%%%%%%%%%%%%%%%%%%%%%%%%%%%%%%%%%%%%

Whereas the cosmogenic neutrino fluxes, discussed above, represent reasonable lower limits on the 
ultrahigh energy neutrino flux, it is also useful to have an upper limit on the 
latter~\cite{Waxman:1999yy,Mannheim:2001wp%
%,Bahcall:2001yr
}. Per construction, the upper
limit from ``visible'' hadronic astrophysical sources, i.\,e. from those sources which are transparent to ultrahigh 
energy cosmic protons
and neutrons, is of the order of the cosmogenic neutrino flux~\cite{Waxman:1999yy,Mannheim:2001wp%
%,Bahcall:2001yr
}\footnote{\label{st79} The old prediction of the cosmogenic neutrino flux from Ref.~\cite{Stecker:1979ah} 
(cf. Fig.~\ref{e2flux_cosmogenic}), which
has been used recently for estimates of black hole detection rates at cosmic ray facilities~\cite{Feng:2001ib,Uehara:2001yk},  
violates the upper limit from ``visible'' hadronic astrophysical sources~\cite{Waxman:1999yy,Mannheim:2001wp%
%,Bahcall:2001yr
} by a factor ${\mathcal O}(2\div 3)$.}.
The upper limit from ``hidden'' hadronic astrophysical sources\footnote{For an early determination of such an upper limit, see 
Ref.~\cite{Berezinsky:1979pd}. Fluxes of this size are predicted in the context of the Z-burst 
scenario~\cite{Fargion:1999ft%
%,Weiler:1999sh,Yoshida:1998it,Fodor:2001qy
} for the highest
energy cosmic rays.}, i.\,e. from those sources from which only 
photons and neutrinos can escape, is much larger~\cite{Mannheim:2001wp} and  
shown in Fig.~\ref{e2flux_cosmogenic}. 
  
\begin{figure}
\begin{center}
%rev
\includegraphics*[bbllx=20pt,bblly=221pt,bburx=585pt,bbury=608pt,%
width=10.8cm]{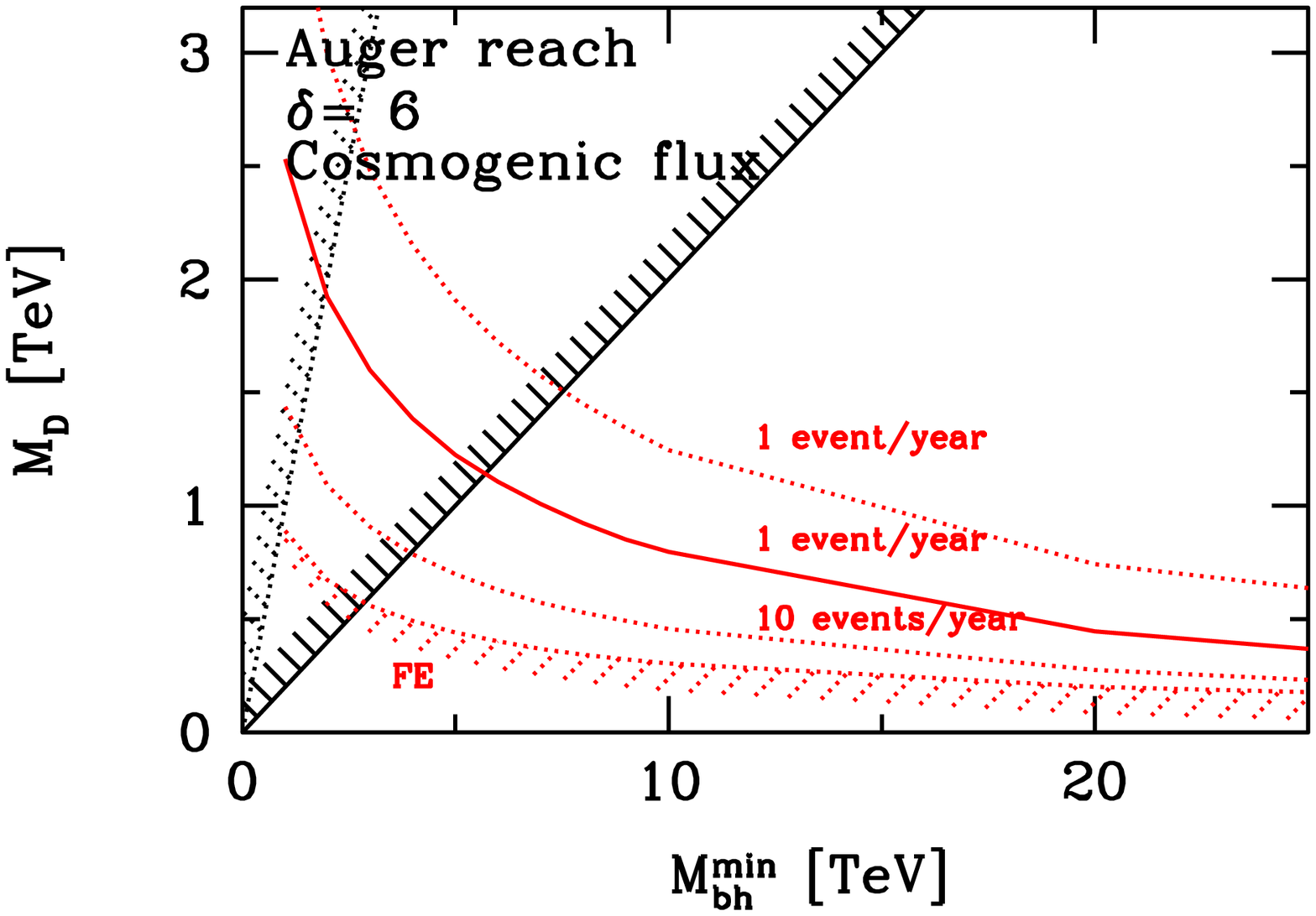}
\includegraphics*[bbllx=20pt,bblly=221pt,bburx=585pt,bbury=608pt,%
width=10.8cm]{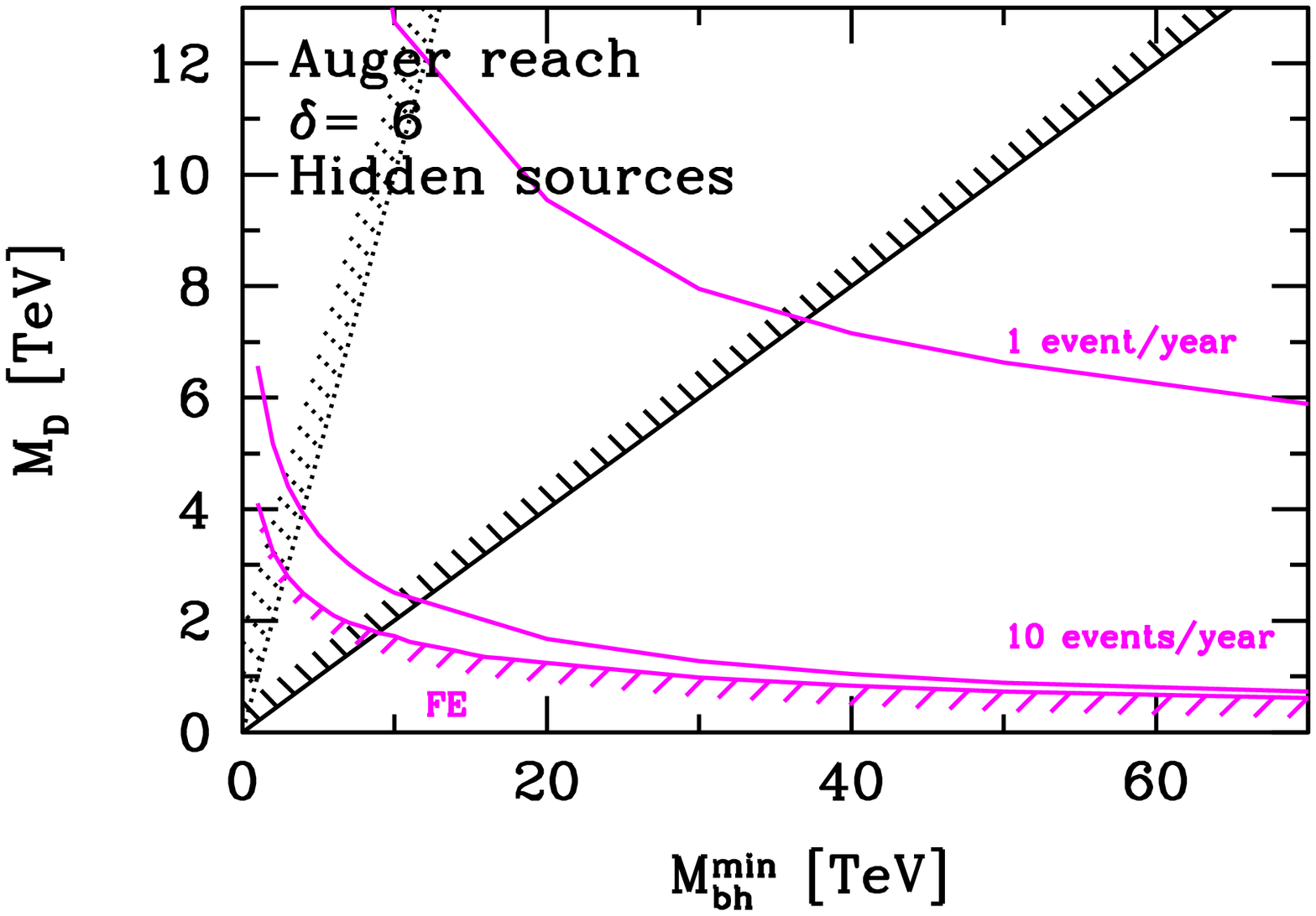}
\caption[dum]{\label{bh_par_auger_d6}
Projected Auger reach in the black hole production parameters for $\delta = 6$ extra dimensions.  
The shaded dotted, $M_D = M_{\rm bh}^{\rm min}$, and shaded solid, 
$M_D=(1/5)\,M_{\rm bh}^{\rm min}$, lines give a rough indication of the 
boundary of applicability of  the semiclassical picture~\cite{Giddings:2001bu}.
{\em Top:} Exploiting the 
cosmogenic neutrino flux from Ref.~\cite{Protheroe:1996ft} (cf. Fig.~\ref{e2flux_cosmogenic}).
The solid (dotted) line(s) assumes a maximum energy $3\cdot 10^{20(21)}$ eV for the ultrahigh energy cosmic
raysis and represents the contour of 1 resp. 10 detected horizontal air shower per year ($10^7$ s) initiated by 
neutrino-nucleon scattering into a black hole with a mass larger than $M_{\rm bh}^{\rm min}$, 
for a fundamental Planck mass $M_D$.
The shaded dotted line labeled ``FE'' indicates the constraint arising from the non-observation of 
horizontal showers by the Fly's Eye collaboration~\cite{Baltrusaitis:1985mt}. 
{\em Bottom:} Exploiting 
the upper limit on the neutrino flux from ``hidden'' hadronic astrophysical sources from
Ref.~\cite{Mannheim:2001wp} (cf. Fig.~\ref{e2flux_cosmogenic}). 
The solid lines represent the contour of 1 and 10 detected horizontal air shower per year ($10^7$ s) initiated by 
neutrino-nucleon scattering into a black hole with a mass larger than $M_{\rm bh}^{\rm min}$, 
for a fundamental Planck mass $M_D$. 
The shaded solid line labeled ``FE'' indicates the constraint arising from the non-observation of 
horizontal showers by the Fly's Eye collaboration~\cite{Baltrusaitis:1985mt}.  
}
\end{center}
\end{figure}

The projected sensitivity of Auger to black hole production can now be investigated, 
for a given neutrino flux, by calculating the event rate~(\ref{rate_auger}). 
Throughout, we shall use the acceptance of the ground array of Auger corresponding to hadronic 
horizontal showers (highest curve in Fig.~4 of Ref.~\cite{Capelle:1998zz}).  
Figure~\ref{bh_par_auger_d6} (top) illustrates the reach in black hole production parameter space, 
for $\delta = 6$ extra dimensions, for the predicted cosmogenic neutrino fluxes from Ref.~\cite{Protheroe:1996ft}.  
We see, that in this case only a handful of events per year can be expected at Auger from the fiducial region of 
parameter space ($M_{\rm bh}^{\rm min}\,\gwig\, 5\,M_D$), with a background from    
``normal'' horizontal air showers initiated by $\nu_e$s and $\bar\nu_e$s 
of about $0.03$ ($0.05$) events per year, for the flux from Ref.~\cite{Protheroe:1996ft} with
$E_{\rm max}=3\cdot 10^{20(21)}$ eV. 
A very similar reach of Auger as the one obtained using the flux labeled $E_{\rm max}=3\cdot 10^{21}$ eV 
in Ref.~\cite{Protheroe:1996ft} 
(dotted line in Fig.~\ref{bh_par_auger_d6} (top)) is obtained using either the flux  
labeled ($m=4$, $z_{\rm max}=4$) in Ref.~\cite{Yoshida:1993pt} (cf. Fig.~\ref{e2flux_cosmogenic})  
or the most recent prediction of the flux from Ref.~\cite{Engel:2001hd}.  
We have refrained from displaying the reach of Auger as obtained from the predicted cosmogenic neutrino flux of 
Ref.~\cite{Stecker:1979ah} (cf. Fig.~\ref{e2flux_cosmogenic}), which
has been used recently for estimates of black hole detection rates at cosmic ray 
facilities~\cite{Feng:2001ib,Uehara:2001yk},
since this prediction is disfavored by more recent 
calculations~\cite{Yoshida:1993pt,Protheroe:1996ft,Yoshida:1997ie,Engel:2001hd}.    

%%%%%%%%%%%%%%%%%%%%%%%%%%%%%%%%FIGURE%%%%%%%%%%%%%%%%%%%%%%%%%%
%%%%%%%%%%%%%%%%%%%%%%%%%%%%%%%%%%%%%%%%%%%%%%%%%%%%%%%%%%%%%%%%%

For a fair comparison of the event rates at Auger with the ones at the LHC, we have to discuss  
how many black hole initiated air showers are needed to discriminate the signal from the Standard Model background.   
Besides the enhanced rate for horizontal showers, what is the characteristic signature of black hole production in
neutrino-induced air showers? In Ref.~\cite{Anchordoqui:2001ei} it was shown that 
these showers may have an ``anomalous'' electromagnetic component: 
about an order of magnitude bigger than Standard Model $\nu_\mu$-initiated showers and at least an order
of magnitude smaller than Standard Model $\nu_e$-initiated showers. It was argued that this represents a very
clean signal and, correspondingly, that a total number of about ${\mathcal O}(20)$  black hole events could give
significant statistics to test this phenomenon.
An inspection of Fig.~\ref{bh_par_auger_d6} (top) leads then to the conclusion that Auger has only a  small
window of opportunity to detect black holes before the start of the LHC, if the ultrahigh energy neutrino
flux is at the level of the cosmogenic one predicted by recent 
calculations~\cite{Yoshida:1993pt,Protheroe:1996ft,Yoshida:1997ie,Engel:2001hd}. 

What is the region of black hole production parameter space which can be probed by Auger under the  most optimistic, but 
still 
conceivable conditions regarding the ultrahigh energy neutrino flux? An answer to this question is 
provided by Fig.~\ref{bh_par_auger_d6} (bottom), which shows the reach in black hole production parameter space, 
for $\delta = 6$ extra dimensions, for the upper limit on the neutrino flux from ``hidden'' 
hadronic astrophysical sources from
Ref.~\cite{Mannheim:2001wp} (cf. Fig.~\ref{e2flux_cosmogenic}). The region of black hole production parameter space 
accessible in this case is impressive and extends much beyond LHC. Note that the Standard Model background is 
here about 3 events/year. If such large ultrahigh energy neutrino fluxes are realized in nature, then the first signs 
of black hole production may be observed at Auger.

%%%%%%%%%%%%%%%%%%%%%%%%%%%%%%%%FIGURE%%%%%%%%%%%%%%%%%%%%%%%%%%
\begin{figure}
\begin{center}
\includegraphics*[bbllx=20pt,bblly=221pt,bburx=570pt,bbury=608pt,%
width=11.cm]{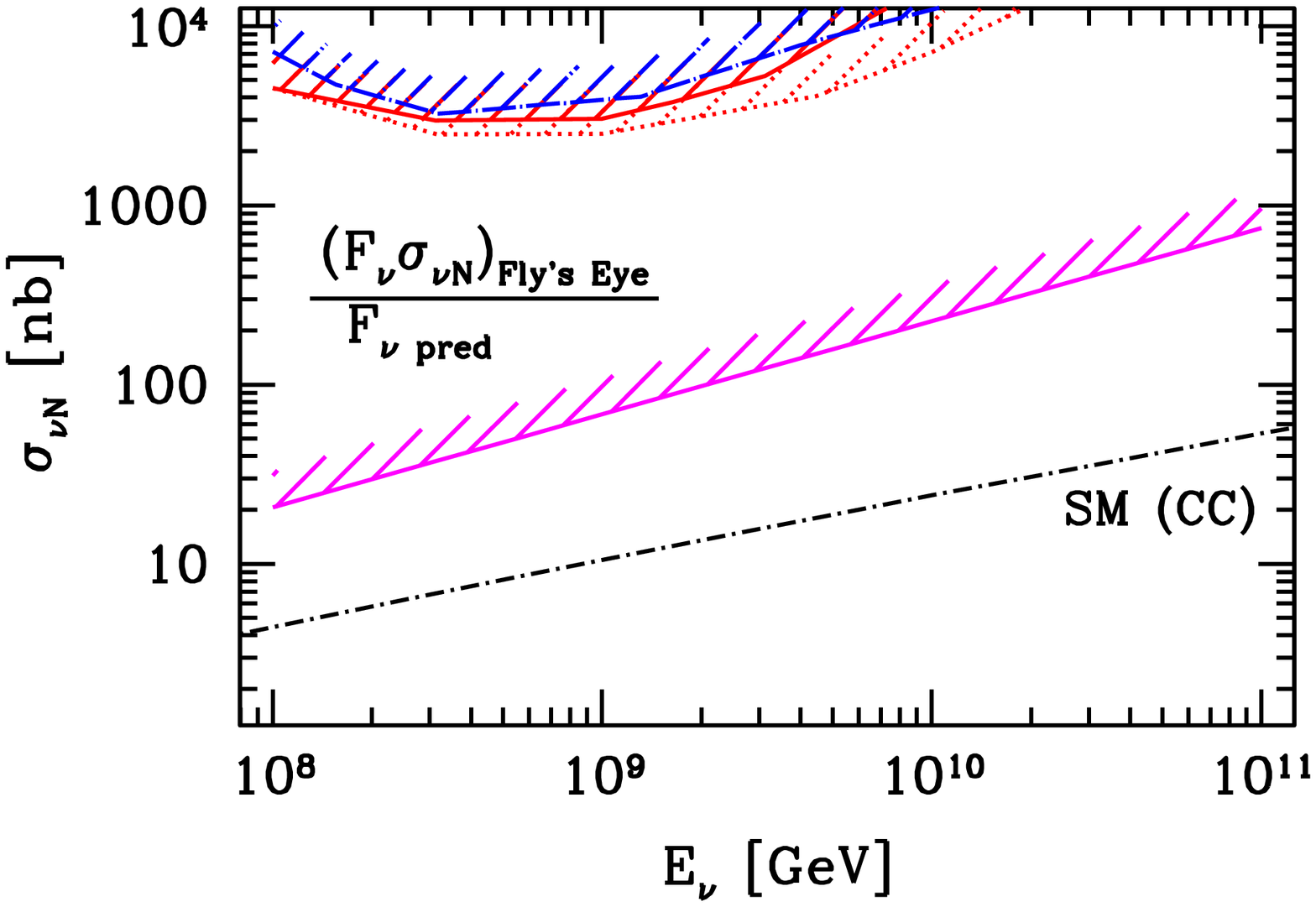}
\caption[dum]{\label{upp_lim_fe}
Upper limit on the neutrino-nucleon cross section obtained from the Fly's Eye limit~(\ref{FlysEyeconstr}), 
for various
predictions of the ultrahigh energy neutrino flux, $F_\nu =\sum_i\left[ F_{\nu_i}+
F_{\bar\nu_i}\right]$ (line styles as in Fig.~\ref{e2flux_cosmogenic}).
Also shown is the Standard Model (SM) charged current (CC) neutrino-nucleon cross section 
(dashed-dotted line). 
}
\end{center}
\end{figure}
%%%%%%%%%%%%%%%%%%%%%%%%%%%%%%%%%%%%%%%%%%%%%%%%%%%%%%%%%%%%%%%%%

%rev
Also shown in Fig.~\ref{bh_par_auger_d6} are the constraints arising from the non-observation of 
horizontal showers by the Fly's Eye collaboration~\cite{Baltrusaitis:1985mt}, 
which are obtained as follows. 
The Fly's Eye collaboration puts upper limits on the product of the total neutrino flux times 
neutrino-nucleon cross section~\cite{Baltrusaitis:1985mt,Reno:1988zf%
%,MacGibbon:1990kk,Morris:1991bb
}
that may be parametrized~\cite{Morris:1991bb} by the 
following power-law, least-squares fit\footnote{Here, again, it is assumed 
that 100\,\% of the incident neutrino energy goes into 
visible, hadronic or electromagnetic shower energy.
Otherwise, one has to take 
into account that the limit applies only for the visible energy~\cite{Tyler:2001gt}.}
\begin{eqnarray}
\label{FlysEyeconstr}
\left( F_\nu\,\sigma_{\nu N}\right) (E_\nu ) 
&\leq & 3.74\cdot 10^{-42}\,\left( \frac{E_\nu}{1\ {\rm GeV}}\right)^{-1.48}
\ 
{\rm GeV}^{-1}\, {\rm s}^{-1}\, {\rm sr}^{-1}
\equiv \left( F_\nu\,\sigma_{\nu N}\right)_{\rm Fly's\ Eye} (E_\nu ) 
\,,
\\[1.5ex] \nonumber
&{\rm for}&\ 10^8\ {\rm GeV}\leq E_\nu\leq 10^{11}\ {\rm GeV}
\hspace{2ex} {\rm and}\hspace{2ex} 
\sigma_{\nu N}(E_\nu ) \leq 10\ \mu{\rm b}\,.
\end{eqnarray}
Thus, for a given, predicted neutrino flux $F_{\nu\,{\rm pred}}$, the Fly's Eye 
constraint~(\ref{FlysEyeconstr}) translates into an upper limit on the 
$\nu N$ cross section, 
$\sigma_{\nu N}(E_\nu )\leq \left( F_\nu\,\sigma_{\nu N}\right)_{\rm Fly's\ Eye}(E_\nu )/
F_{\nu\,{\rm pred}(E_\nu )}$~\cite{Morris:1994wg,Tyler:2001gt} 
(for an early reasoning along these lines using older data, see Ref.~\cite{Berezinsky:1974kz}), which is shown
in Fig.~\ref{upp_lim_fe}, for various flux predictions from Fig.~\ref{e2flux_cosmogenic}. 
Finally, a comparison of the prediction $\sigma_{\nu N}=\sigma_{\nu N}^{\rm SM}+\sigma_{\nu N}^{\rm bh}$,
where $\sigma_{\nu N}$ is the Standard Model contribution, 
with the upper limits of Fig.~\ref{upp_lim_fe} yields then excluded regions in
black hole production parameter space, as 
%rev
those shown in Fig.~\ref{bh_par_auger_d6}. 

%rev
From an inspection of Fig.~\ref{bh_par_auger_d6} (bottom) one finds that the Fly's Eye constraints
on black hole production, although more model dependent, compare favourably with the currently available limits 
on TeV-scale gravity~\cite{Peskin:2000ti%
%,Abe:2001nq
}, at least as long as a neutrino flux on the level of the upper limit from
``hidden'' hadronic sources is realized in nature. In the case of the more conservative cosmogenic neutrino flux, however,  
the Fly's Eye constraints are only marginally competitive with the above mentioned limits (cf. Fig.~\ref{bh_par_auger_d6} 
(top)).

%\section{Conclusions}
{\em 4.} 
We considered the reach of the LHC and the Pierre Auger Observatory to black hole production in the
context of extra dimension scenarios with TeV-scale gravity. 
%rev
Moreover, we have also derived constraints   
in the black hole production parameter space from the non-observation of horizontal showers by the Fly's Eye collaboration.
We found that if the ultrahigh energy neutrino
flux is at the (almost guaranteed) level of the cosmogenic one predicted by recent 
calculations~\cite{Yoshida:1993pt,Protheroe:1996ft,Yoshida:1997ie,Engel:2001hd}, Auger has only a small
window of opportunity before the start of the LHC to observe the first signs of black hole production in horizontal air 
showers
initiated by ultrahigh energy neutrinos. 
If, on the other hand, larger ultrahigh energy neutrino fluxes on the level of the upper limit from 
``hidden'' hadronic astrophysical sources~\cite{Mannheim:2001wp} are realized in nature, then the first signs 
of black hole production may be observed at Auger.
%rev
Moreover, in this case, the Fly's Eye constraints, although more model-dependent, turn out to be competitive with other 
currently available constraints on TeV-scale gravity which are mainly based on interactions associated with
Kaluza-Klein gravitons.
It remains to be seen from a full simulation whether the characteristics of 
Standard Model and black hole initiated air showers are sufficiently distinctive~\cite{Anchordoqui:2001ei} 
to successfully attribute an eventual excess in events to black hole production rather than to an enhancement in the 
ultrahigh energy neutrino flux.
From the experience with the phenomenology of sphaleron production~\cite{Morris:1994wg} 
we think it is worthwhile
to work out also the signature expected to be seen in neutrino telescopes such as 
AMANDA/ICECUBE~\cite{Halzen:1999jy} and
RICE~\cite{Frichter:1999kr}, which might offer additional ways to look for black holes before the start of the LHC.   

\section*{Acknowledgements}

We thank 
A. Brandenburg, S. Huber, M. Kowalski, K. Mannheim, F. Schrempp, E. Waxman, J. Wells, 
and E. Zas for fruitful discussions.


\begin{thebibliography}{99}

%Black hole production in 4d

%rev
%\cite{'tHooft:1987rb}
\bibitem{'tHooft:1987rb}
G.~'t Hooft,
%``Graviton Dominance In Ultrahigh-Energy Scattering,''
Phys.\ Lett.\ B {198} (1987) 61; 
%%CITATION = PHLTA,B198,61;%%
%
%\cite{'tHooft:1988wk}
%\bibitem{'tHooft:1988wk}
%G.~'t Hooft,
%``On The Factorization Of Universal Poles In A Theory Of Gravitating Point Particles,''
Nucl.\ Phys.\ B {304} (1988) 867;
%%CITATION = NUPHA,B304,867;%%
%
%\cite{Amati:1987wq}
%\bibitem{Amati:1987wq}
D.~Amati, M.~Ciafaloni and G.~Veneziano,
%``Superstring Collisions At Planckian Energies,''
Phys.\ Lett.\ B {197} (1987) 81; 
%
%%CITATION = PHLTA,B197,81;%%
%\cite{Amati:1988uf}
%\bibitem{Amati:1988uf}
%D.~Amati, M.~Ciafaloni and G.~Veneziano,
%``Classical And Quantum Gravity Effects From Planckian Energy Superstring Collisions,''
Int.\ J.\ Mod.\ Phys.\ A {3} (1988) 1615; 
%%CITATION = IMPAE,A3,1615;%%
%
%\cite{Amati:1989tn}
%\bibitem{Amati:1989tn}
%D.~Amati, M.~Ciafaloni and G.~Veneziano,
%``Can Space-Time Be Probed Below The String Size?,''
Phys.\ Lett.\ B {216} (1989) 41; 
%%CITATION = PHLTA,B216,41;%%
%
%\cite{Amati:1990xe}
%\bibitem{Amati:1990xe}
%D.~Amati, M.~Ciafaloni and G.~Veneziano,
%``Higher Order Gravitational Deflection And Soft Bremsstrahlung In Planckian Energy Superstring Collisions,''
Nucl.\ Phys.\ B {347} (1990) 550; 
%%CITATION = NUPHA,B347,550;%%
%
%\cite{Amati:1992zb}
%\bibitem{Amati:1992zb}
%D.~Amati, M.~Ciafaloni and G.~Veneziano,
%``Planckian scattering beyond the semiclassical approximation,''
Phys.\ Lett.\ B {289} (1992) 87;
%%CITATION = PHLTA,B289,87;%%
%
%\cite{Amati:1993tb}
%\bibitem{Amati:1993tb}
%D.~Amati, M.~Ciafaloni and G.~Veneziano,
%``Effective action and all order gravitational eikonal at Planckian energies,''
Nucl.\ Phys.\ B {403} (1993) 707;
%%CITATION = NUPHA,B403,707;%%
%
%\cite{Aref'eva:1995qs}
%\bibitem{Aref'eva:1995qs}
I.~Y.~Aref'eva, K.~S.~Viswanathan and I.~V.~Volovich,
%``Planckian energy scattering, colliding plane gravitational waves and black hole creation,''
Nucl.\ Phys.\ B  {452} (1995) 346
[Erratum-ibid.\ B  {462} (1995) 613].
%[arXiv:hep-th/9412157].
%%CITATION = HEP-TH 9412157;%%

%Large extra dimensions and low scale QG

%\cite{Arkani-Hamed:1998rs}
\bibitem{Arkani-Hamed:1998rs}
N.~Arkani-Hamed, S.~Dimopoulos and G.~R.~Dvali,
%``The hierarchy problem and new dimensions at a millimeter,''
Phys.\ Lett.\ B { 429} (1998) 263;
%[hep-ph/9803315].
%%CITATION = HEP-PH 9803315;%%
%
%\cite{Antoniadis:1998ig}
%\bibitem{Antoniadis:1998ig}
I.~Antoniadis, N.~Arkani-Hamed, S.~Dimopoulos and G.~R.~Dvali,
%``New dimensions at a millimeter to a Fermi and superstrings at a TeV,''
Phys.\ Lett.\ B { 436} (1998) 257;
%[hep-ph/9804398].
%%CITATION = HEP-PH 9804398;%%
%
%\cite{Arkani-Hamed:1999nn}
%\bibitem{Arkani-Hamed:1999nn}
N.~Arkani-Hamed, S.~Dimopoulos and G.~R.~Dvali,
%``Phenomenology, astrophysics and cosmology of theories with  sub-millimeter dimensions and TeV scale quantum gravity,''
Phys.\ Rev.\ D { 59} (1999) 086004.
%[hep-ph/9807344].
%%CITATION = HEP-PH 9807344;%%

%noncompact extra dimensions and TeV scale gravity

%\cite{Randall:1999ee}
\bibitem{Randall:1999ee}
L.~J.~Randall and R.~Sundrum,
%``A large mass hierarchy from a small extra dimension,''
Phys.\ Rev.\ Lett.\  {83} (1999) 3370.
%[arXiv:hep-ph/9905221].
%%CITATION = HEP-PH 9905221;%%


%Black hole production

%\cite{Argyres:1998qn}
\bibitem{Argyres:1998qn}
P.~C.~Argyres, S.~Dimopoulos and J.~March-Russell,
%``Black holes and sub-millimeter dimensions,''
Phys.\ Lett.\ B {441} (1998) 96; 
%[hep-th/9808138].
%%CITATION = HEP-TH 9808138;%%
%
%\cite{Banks:1999gd}
%\bibitem{Banks:1999gd}
T.~Banks and W.~Fischler,
%``A model for high energy scattering in quantum gravity,''
hep-th/9906038; 
%%CITATION = HEP-TH 9906038;%%
%
%\cite{Aref'eva:1999bm}
%\bibitem{Aref'eva:1999bm}
I.~Y.~Aref'eva,
%``High energy scattering in the brane-world and black hole production,''
%arXiv:
hep-th/9910269.
%%CITATION = HEP-TH 9910269;%%

%\cite{Emparan:2000rs}
\bibitem{Emparan:2000rs}
R.~Emparan, G.~T.~Horowitz and R.~C.~Myers,
%``Black holes radiate mainly on the brane,''
Phys.\ Rev.\ Lett.\  {85} (2000) 499.
%[hep-th/0003118].
%%CITATION = HEP-TH 0003118;%%

%\cite{Giddings:2000ay}
\bibitem{Giddings:2000ay}
S.~B.~Giddings and E.~Katz,
%``Effective theories and black hole production in warped  compactifications,''
%arXiv:
hep-th/0009176;
%%CITATION = HEP-TH 0009176;%%
%
%\cite{Emparan:2001ce}
%\bibitem{Emparan:2001ce}
R.~Emparan,
%``Exact gravitational shockwaves and Planckian scattering on branes,''
Phys.\ Rev.\ D {64} (2001) 024025.
%[arXiv:hep-th/0104009].
%%CITATION = HEP-TH 0104009;%%

%Black hole production at the LHC

%\cite{Giddings:2001bu}
\bibitem{Giddings:2001bu}
S.~B.~Giddings and S.~Thomas,
%``High energy colliders as black hole factories: The end of short  distance physics,''
hep-ph/0106219.
%%CITATION = HEP-PH 0106219;%%


%\cite{Dimopoulos:2001hw}
\bibitem{Dimopoulos:2001hw}
S.~Dimopoulos and G.~Landsberg,
%``Black holes at the LHC,''
Phys.\ Rev.\ Lett.\  {87} (2001) 161602.
%[arXiv:hep-ph/0106295].
%%CITATION = HEP-PH 0106295;%%

%LHC

%\cite{Evans:2001mn}
\bibitem{Evans:2001mn}
L.~R.~Evans,
%``The Large Hadron Collider: Present status and prospects,''
CERN-OPEN-2001-027, presented at: 
18th International Conference on High Energy Accelerators, Tsukuba, Japan, 26 - 30 Mar 2001.

%Black hole production at LHC continued


%\cite{Dimopoulos:2001qe}
\bibitem{Dimopoulos:2001qe}
S.~Dimopoulos and R.~Emparan,
%``String balls at the LHC and beyond,''
hep-ph/0108060.
%%CITATION = HEP-PH 0108060;%%

%\cite{Hossenfelder:2001dn}
\bibitem{Hossenfelder:2001dn}
S.~Hossenfelder, S.~Hofmann, M.~Bleicher and H.~St\"ocker,
%``Quasi-Stable Black Holes at LHC,''
hep-ph/0109085;
%%CITATION = HEP-PH 0109085;%%
%
%\cite{Giddings:2001ih}
%\bibitem{Giddings:2001ih}
S.~B.~Giddings,
%``Black hole production in TeV-scale gravity, and the future of high  energy physics,''
%arXiv:
hep-ph/0110127;
%%CITATION = HEP-PH 0110127;%%
%
%\cite{Cheung:2001ue}
%\bibitem{Cheung:2001ue}
K.~Cheung,
%``Black hole production and large extra dimensions,''
%arXiv:
hep-ph/0110163;
%%CITATION = HEP-PH 0110163;%%
%
%rev
%\cite{Casadio:2001wh}
%\bibitem{Casadio:2001wh}
R.~Casadio and B.~Harms,
%``Can black holes and naked singularities be detected in accelerators?,''
%arXiv:
hep-th/0110255.
%%CITATION = HEP-TH 0110255;%%

%Black holes in nuN scattering and cosmic rays

%\cite{Feng:2001ib}
\bibitem{Feng:2001ib}
J.~L.~Feng and A.~D.~Shapere,
%``Black Hole Production by Cosmic Rays,''
hep-ph/0109106.
%%CITATION = HEP-PH 0109106;%%

%Auger Observatory

%\cite{Zavrtanik:2000zi}
\bibitem{Zavrtanik:2000zi}
D.~Zavrtanik  [AUGER Collaboration],
%``The Pierre Auger Observatory,''
Nucl.\ Phys.\ Proc.\ Suppl.\  {85} (2000) 324; 
%%CITATION = NUPHZ,85,324;%%
%
%\cite{Zepeda:2000zk}
%\bibitem{Zepeda:2000zk}
A.~Zepeda,
%``Ultra high energy cosmic rays and the Pierre Auger project,''.
in: E. Nardi (Ed.),  
3rd Latin American Symposium On High-Energy Physics, 
2-8 Apr 2000, Cartagena de Indias, Colombia,  Bristol, IOP, 2000.
%http://jhep.sissa.it/archive/prhep/preproceeding/005/040/silafae-Z5.pdf

%Further disucssion: Black holes in nuN scattering and cosmic rays

%\cite{Anchordoqui:2001ei}
\bibitem{Anchordoqui:2001ei}
L.~Anchordoqui and H.~Goldberg,
%``Experimental signature for black hole production in neutrino air  showers,''
%arXiv:
hep-ph/0109242.
%%CITATION = HEP-PH 0109242;%%

%\cite{Emparan:2001kf}
\bibitem{Emparan:2001kf}
R.~Emparan, M.~Masip and R.~Rattazzi,
%``Cosmic rays as probes of large extra dimensions and TeV gravity,''
%arXiv:
hep-ph/0109287.
%%CITATION = HEP-PH 0109287;%%

%black holes in ICECUBE

%\cite{Uehara:2001yk}
\bibitem{Uehara:2001yk}
Y.~Uehara,
%``Production and Detection of Black Holes at Neutrino Array,''
%arXiv:
hep-ph/0110382.
%%CITATION = HEP-PH 0110382;%%

%ICECUBE

%\cite{Halzen:1999jy}
\bibitem{Halzen:1999jy}
F.~Halzen {\it et al.}  [AMANDA Collaboration],
%``From the first neutrino telescope, the antarctic muon and neutrino  detector array AMANDA, to the IceCube observatory,''
in: B.L. Dingus, D.B. Kieda, M.H. Salamon (Eds.),  
26th International Cosmic Ray Conference (ICRC 99), Salt Lake City, UT, 17-25 Aug 1999,
Melville, AIP, 2000, pp. 428-431.

%Fly's Eye limits

%\cite{Baltrusaitis:1985mt}
\bibitem{Baltrusaitis:1985mt}
R.~M.~Baltrusaitis {\it et al.},
%``Limits On Deeply Penetrating Particles In The > 10**17-Ev Cosmic Ray Flux,''
Phys.\ Rev.\ D {31} (1985) 2192.
%%CITATION = PHRVA,D31,2192;%%

%Fly's Eye 

%\cite{Baltrusaitis:1985mx}
\bibitem{Baltrusaitis:1985mx}
R.~M.~Baltrusaitis {\it et al.},
%``The Utah Fly's Eye Detector,''
Nucl.\ Instrum.\ Meth.\ A {240} (1985) 410.
%%CITATION = NUIMA,A240,410;%%



%TeV-scale gravity and contact interactions from KK exchange

%collider

%\cite{Giudice:1999ck}
\bibitem{Giudice:1999ck}
G.~F.~Giudice, R.~Rattazzi and J.~D.~Wells,
%``Quantum gravity and extra dimensions at high-energy colliders,''
Nucl.\ Phys.\ B {544} (1999) 3
and erratum to appear [hep-ph/9811291 v2].
%%CITATION = HEP-PH 9811291;%%

%\cite{Mirabelli:1999rt}
\bibitem{Mirabelli:1999rt}
E.~A.~Mirabelli, M.~Perelstein and M.~E.~Peskin,
%``Collider signatures of new large space dimensions,''
Phys.\ Rev.\ Lett.\  {82} (1999) 2236;
%[hep-ph/9811337].
%%CITATION = HEP-PH 9811337;%%
%
%\cite{Han:1999sg}
%\bibitem{Han:1999sg}
T.~Han, J.~D.~Lykken and R.~Zhang,
%``On Kaluza-Klein states from large extra dimensions,''
Phys.\ Rev.\ D {59} (1999) 105006;
%[hep-ph/9811350].
%%CITATION = HEP-PH 9811350;%%
%
%\cite{Hewett:1999sn}
%\bibitem{Hewett:1999sn}
J.~L.~Hewett,
%``Indirect collider signals for extra dimensions,''
Phys.\ Rev.\ Lett.\  {82} (1999) 4765; 
%[hep-ph/9811356].
%%CITATION = HEP-PH 9811356;%%
%
%\cite{Mathews:1999kf}
%\bibitem{Mathews:1999kf}
P.~Mathews, S.~Raychaudhuri and K.~Sridhar,
%``Getting to the top with extra dimensions,''
Phys.\ Lett.\ B {450} (1999) 343;
%[hep-ph/9811501].
%%CITATION = HEP-PH 9811501;%%
%
%\cite{Rizzo:1999fm}
%\bibitem{Rizzo:1999fm}
T.~G.~Rizzo,
%``More and more indirect signals for extra dimensions at more and more  colliders,''
Phys.\ Rev.\ D {59} (1999) 115010; 
%[hep-ph/9901209].
%%CITATION = HEP-PH 9901209;%%
%
%\cite{Mathews:1999qn}
%\bibitem{Mathews:1999qn}
P.~Mathews, S.~Raychaudhuri and K.~Sridhar,
%``Large extra dimensions and deep-inelastic scattering at HERA,''
Phys.\ Lett.\ B {455} (1999) 115;
%[hep-ph/9812486].
%%CITATION = HEP-PH 9812486;%%
%
%\cite{Cheung:1999qh}
%\bibitem{Cheung:1999qh}
K.~Cheung,
%``Global lepton quark neutral current constraint on low scale gravity  model,''
Phys.\ Lett.\ B {460} (1999) 383;
%[hep-ph/9904510].
%%CITATION = HEP-PH 9904510;%%
%
%\cite{Adloff:2000dp}
%\bibitem{Adloff:2000dp}
C.~Adloff {\it et al.}  [H1 Collaboration],
%``Search for compositeness, leptoquarks and large extra dimensions in e q  contact interactions at HERA,''
Phys.\ Lett.\ B {479} (2000) 358;
%[hep-ex/0003002].
%%CITATION = HEP-EX 0003002;%%
%
%collider constraints on warped extra dimension secenario
%
%rev
%\cite{Davoudiasl:2000jd}
%\bibitem{Davoudiasl:2000jd}
H.~Davoudiasl, J.~L.~Hewett and T.~G.~Rizzo,
%``Phenomenology of the Randall-Sundrum gauge hierarchy model,''
Phys.\ Rev.\ Lett.\  {84} (2000) 2080.
%[arXiv:hep-ph/9909255].
%%CITATION = HEP-PH 9909255;%%

%astrophysical constraints on add scenario

\bibitem{Cullen:1999hc}
S.~Cullen and M.~Perelstein,
%``SN1987A constraints on large compact dimensions,''
Phys.\ Rev.\ Lett.\  {83} (1999) 268;
%[arXiv:hep-ph/9903422].
%%CITATION = HEP-PH 9903422;%%
%
%\cite{Barger:1999jf}
%\bibitem{Barger:1999jf}
V.~Barger, T.~Han, C.~Kao and R.~J.~Zhang,
%``Astrophysical constraints on large extra dimensions,''
Phys.\ Lett.\ B {461} (1999) 34;
%[arXiv:hep-ph/9905474].
%%CITATION = HEP-PH 9905474;%%
%
%\cite{Hanhart:2001er}
%\bibitem{Hanhart:2001er}
C.~Hanhart, D.~R.~Phillips, S.~Reddy and M.~J.~Savage,
%``Extra dimensions, SN1987a, and nucleon nucleon scattering data,''
Nucl.\ Phys.\ B {595} (2001) 335; 
%[arXiv:nucl-th/0007016].
%%CITATION = NUCL-TH 0007016;%%
%
%\cite{Hanhart:2001fx}
%\bibitem{Hanhart:2001fx}
C.~Hanhart, J.~A.~Pons, D.~R.~Phillips and S.~Reddy,
%``The likelihood of GODs' existence: Improving the SN 1987a constraint on  the size of large compact dimensions,''
Phys.\ Lett.\ B {509} (2001) 1;
%[arXiv:astro-ph/0102063].
%%CITATION = ASTRO-PH 0102063;%%
%
%\cite{Hannestad:2001jv}
%\bibitem{Hannestad:2001jv}
S.~Hannestad and G.~Raffelt,
%``New supernova limit on large extra dimensions,''
Phys.\ Rev.\ Lett.\  {87} (2001) 051301;
%[arXiv:hep-ph/0103201].
%%CITATION = HEP-PH 0103201;%%
%
%\cite{Hannestad:2001xi}
%\bibitem{Hannestad:2001xi}
%S.~Hannestad and G.~G.~Raffelt,
%``Stringent neutron-star limits on large extra dimensions,''
%arXiv:
hep-ph/0110067.
%%CITATION = HEP-PH 0110067;%%

%cosmological constraints on add scenario

%\cite{Hall:1999mk}
\bibitem{Hall:1999mk}
L.~J.~Hall and D.~R.~Smith,
%``Cosmological constraints on theories with large extra dimensions,''
Phys.\ Rev.\ D {60} (1999) 085008; 
%[arXiv:hep-ph/9904267].
%%CITATION = HEP-PH 9904267;%%
%
%rev 
%\cite{Fairbairn:2001ct}
%\bibitem{Fairbairn:2001ct}
M.~Fairbairn,
%``Cosmological constraints on large extra dimensions,''
Phys.\ Lett.\ B {508} (2001) 335;
%[arXiv:hep-ph/0101131].
%%CITATION = HEP-PH 0101131;%%
%
%\cite{Hannestad:2001nq}
%\bibitem{Hannestad:2001nq}
S.~Hannestad,
%``Strong constraint on large extra dimensions from cosmology,''
Phys.\ Rev.\ D {64} (2001) 023515.
%[arXiv:hep-ph/0102290].
%%CITATION = HEP-PH 0102290;%%


%cosmic ray

%\cite{Nussinov:1999jt}
\bibitem{Nussinov:1999jt}
S.~Nussinov and R.~Shrock,
%``Some remarks on theories with large compact dimensions and TeV-scale  quantum gravity,''
Phys.\ Rev.\ D {59} (1999) 105002;
%[hep-ph/9811323].
%%CITATION = HEP-PH 9811323;%%
%
%\cite{Jain:2000pu}
%\bibitem{Jain:2000pu}
P.~Jain, D.~W.~McKay, S.~Panda and J.~P.~Ralston,
%``Extra dimensions and strong neutrino nucleon interactions above  10**19-eV: Breaking the GZK barrier,''
Phys.\ Lett.\ B {484} (2000) 267.
%[hep-ph/0001031].
%%CITATION = HEP-PH 0001031;%%

%\cite{Tyler:2001gt}
\bibitem{Tyler:2001gt}
C.~Tyler, A.~V.~Olinto and G.~Sigl,
%``Cosmic neutrinos and new physics beyond the electroweak scale,''
Phys.\ Rev.\ D {63} (2001) 055001.
%[hep-ph/0002257].
%%CITATION = HEP-PH 0002257;%%


%\cite{Kachelriess:2000cb}
\bibitem{Kachelriess:2000cb}
M.~Kachelriess and M.~Pl\"umacher,
%``Ultrahigh energy neutrino interactions and weak-scale string theories,''
Phys.\ Rev.\ D {62} (2000) 103006;
%[astro-ph/0005309].
%%CITATION = ASTRO-PH 0005309;%%
%
%\cite{Anchordoqui:2001mk}
%\bibitem{Anchordoqui:2001mk}
L.~A.~Anchordoqui, T.~P.~McCauley, S.~Reucroft and J.~Swain,
%``Echoes of the fifth dimension?,''
Phys.\ Rev.\ D {63} (2001) 027303; 
%[hep-ph/0009319].
%%CITATION = HEP-PH 0009319;%%
%
%\cite{Anchordoqui:2001uh}
%\bibitem{Anchordoqui:2001uh}
L.~Anchordoqui, H.~Goldberg, T.~McCauley, T.~Paul, S.~Reucroft and J.~Swain,
%``Extensive air showers with TeV-scale quantum gravity,''
Phys.\ Rev.\ D {63} (2001) 124009;
%[hep-ph/0011097].
%%CITATION = HEP-PH 0011097;%%
%
%\cite{Alvarez-Muniz:2001mk}
%\bibitem{Alvarez-Muniz:2001mk}
J.~Alvarez-Muniz, F.~Halzen, T.~Han and D.~Hooper,
%``Phenomenology of high energy neutrinos in low-scale quantum gravity  models,''
hep-ph/0107057;
%%CITATION = HEP-PH 0107057;%%
%
%\cite{Kachelriess:2001jq}
%\bibitem{Kachelriess:2001jq}
M.~Kachelriess and M.~Pl\"umacher,
%``Remarks on the high-energy behaviour of cross-sections in weak-scale string theories,''
%arXiv:
hep-ph/0109184.
%%CITATION = HEP-PH 0109184;%%

%Summary of constraints on large extra dimensions

%\cite{Peskin:2000ti}
\bibitem{Peskin:2000ti}
M.~E.~Peskin,
%``Theoretical summary,''
%arXiv:
hep-ph/0002041;
%%CITATION = HEP-PH 0002041;%%
%
%\cite{Abe:2001nq}
%\bibitem{Abe:2001nq}
T.~Abe {\it et al.}  [American Linear Collider Working Group Collaboration],
%``Linear collider physics resource book for Snowmass 2001. 3: Studies of  exotic and standard model physics,''
%arXiv:
hep-ex/0106057;
%%CITATION = HEP-EX 0106057;%%
%
%rev
%\cite{Pagliarone:2001ff}
%\bibitem{Pagliarone:2001ff}
C.~Pagliarone,
%``Extra dimensions and black hole production,''
%arXiv:
hep-ex/0111063.
%%CITATION = HEP-EX 0111063;%%


%Black hole Schwarzschild radius

%\cite{Myers:1986un}
\bibitem{Myers:1986un}
R.~C.~Myers and M.~J.~Perry,
%``Black Holes In Higher Dimensional Space-Times,''
Annals Phys.\  {172} (1986) 304.
%%CITATION = APNYA,172,304;%%

%suppression of not suppression?

%\cite{Voloshin:2001vs}
\bibitem{Voloshin:2001vs}
M.~B.~Voloshin,
%``Semiclassical suppression of black hole production in particle  collisions,''
Phys.\ Lett.\ B {518} (2001) 137;
%[arXiv:hep-ph/0107119].
%%CITATION = HEP-PH 0107119;%%
%rev
%\cite{Voloshin:2001fe}
%\bibitem{Voloshin:2001fe}
%M.~B.~Voloshin,
%``More remarks on suppression of large black hole production in particle  collisions,''
%arXiv:
hep-ph/0111099.
%%CITATION = HEP-PH 0111099;%%

%production of semiclassical field configurations

%\cite{Klinkhamer:1984di}
\bibitem{Klinkhamer:1984di}
F.~R.~Klinkhamer and N.~S.~Manton,
%``A Saddle Point Solution In The Weinberg-Salam Theory,''
Phys.\ Rev.\ D {30} (1984) 2212.
%%CITATION = PHRVA,D30,2212;%%

%\cite{Ringwald:1990ee}
\bibitem{Ringwald:1990ee}
A.~Ringwald,
%``High-Energy Breakdown Of Perturbation Theory In The Electroweak Instanton Sector,''
Nucl.\ Phys.\ B {330} (1990) 1;
%%CITATION = NUPHA,B330,1;%%
%
%\cite{Espinosa:1990qn}
%\bibitem{Espinosa:1990qn}
O.~Espinosa,
%``High-Energy Behavior Of Baryon And Lepton Number Violating Scattering Amplitudes And 
%Breakdown Of Unitarity In The Standard Model,''
Nucl.\ Phys.\ B {343} (1990) 310;
%%CITATION = NUPHA,B343,310;%%
%
%\cite{Farrar:1990vb}
%\bibitem{Farrar:1990vb}
G.~R.~Farrar and R.~Meng,
%``Baryon Number Violation In High-Energy Collisions,''
Phys.\ Rev.\ Lett.\  {65} (1990) 3377.
%%CITATION = PRLTA,65,3377;%%
%
%\cite{Ringwald:1991bh}
\bibitem{Ringwald:1991bh}
A.~Ringwald and C.~Wetterich,
%``How strong are weak interactions in the multi - TeV range?,''
Nucl.\ Phys.\ B {353} (1991) 303.
%%CITATION = NUPHA,B353,303;%%
%
%\cite{Ringwald:1991qz}
\bibitem{Ringwald:1991qz}
A.~Ringwald, F.~Schrempp and C.~Wetterich,
%``Phenomenology of geometrical flavor interactions at TeV energies,''
Nucl.\ Phys.\ B {365} (1991) 3.
%%CITATION = NUPHA,B365,3;%%

%Collider Phenomenology of sphaleron production and decsy

%\cite{Morris:1991bb}
\bibitem{Morris:1991bb}
D.~A.~Morris and R.~Rosenfeld,
%``Muon bundles from cosmic ray multi - W phenomena,''
Phys.\ Rev.\ D {44} (1991) 3530.
%%CITATION = PHRVA,D44,3530;%%

%constraints for new physics from Fly's Eye + cosmogenic flux

%\cite{Morris:1994wg}
\bibitem{Morris:1994wg}
D.~A.~Morris and A.~Ringwald,
%``Cosmic ray signatures of multi - W processes,''
Astropart.\ Phys.\  {2} (1994) 43.
%[hep-ph/9308269].
%%CITATION = HEP-PH 9308269;%%


%\cite{Gibbs:1995cw}
\bibitem{Gibbs:1995cw}
M.~J.~Gibbs, A.~Ringwald, B.~R.~Webber and J.~T.~Zadrozny,
%``Monte Carlo simulation of baryon and lepton number violating processes at high-energies,''
Z.\ Phys.\ C {66} (1995) 285.
%[hep-ph/9406266].
%%CITATION = HEP-PH 9406266;%%

%\cite{Mattis:1992bj}
\bibitem{Mattis:1992bj}
M.~P.~Mattis,
%``The Riddle of high-energy baryon number violation,''
Phys.\ Rept.\  {214} (1992) 159;
%%CITATION = PRPLC,214,159;%%
%
%\cite{Tinyakov:1993dr}
%\bibitem{Tinyakov:1993dr}
P.~G.~Tinyakov,
%``Instanton like transitions in high-energy collisions,''
Int.\ J.\ Mod.\ Phys.\ A {8} (1993) 1823;
%%CITATION = IMPAE,A8,1823;%%
%
%\cite{Voloshin:1994yp}
%\bibitem{Voloshin:1994yp}
M.~B.~Voloshin, 
%``Nonperturbative methods,''
hep-ph/9409344;
%%CITATION = HEP-PH 9409344;%%
%
%\cite{Rubakov:1996vz}
%\bibitem{Rubakov:1996vz}
V.~A.~Rubakov and M.~E.~Shaposhnikov,
%``Electroweak baryon number non-conservation in the early universe and in  high-energy collisions,''
Usp.\ Fiz.\ Nauk {166} (1996) 493
[Phys.\ Usp.\  {39} (1996) 461].
%[hep-ph/9603208].
%%CITATION = HEP-PH 9603208;%%

%pflib

%\cite{Plothow-Besch:1993qj}
\bibitem{Plothow-Besch:1993qj}
H.~Plothow-Besch,
%``PDFLIB: A Library of all available parton density functions of the nucleon, 
%the pion and the photon and the corresponding alpha-s calculations,''
Comput.\ Phys.\ Commun.\  {75} (1993) 396;
%%CITATION = CPHCB,75,396;%%
%
%\cite{Plothow-Besch:1995ci}
%\bibitem{Plothow-Besch:1995ci}
%H.~Plothow-Besch,
%``The Parton distribution function library,''
Int.\ J.\ Mod.\ Phys.\ A {10} (1995) 2901; 
%%CITATION = IMPAE,A10,2901;%%
%
%\cite{Plothow-Besch:1995ci}
%\bibitem{Plothow-Besch:2001}
%H.~Plothow-Besch, 
http://consult.cern.ch/writeups/pdflib/main.ps

%pdf sets

%\cite{Lai:2000wy}
\bibitem{Lai:2000wy}
H.~L.~Lai {\it et al.}  [CTEQ Collaboration],
%``Global {QCD} analysis of parton structure of the nucleon: CTEQ5 parton  distributions,''
Eur.\ Phys.\ J.\ C {12} (2000) 375.
%[arXiv:hep-ph/9903282].
%%CITATION = HEP-PH 9903282;%%

%Hawking radiation

%\cite{Hawking:1975sw}
\bibitem{Hawking:1975sw}
S.~W.~Hawking,
%``Particle Creation By Black Holes,''
Commun.\ Math.\ Phys.\  {43} (1975) 199.
%%CITATION = CMPHA,43,199;%%

%cteq3 pdf

%\cite{Lai:1995bb}
\bibitem{Lai:1995bb}
H.~L.~Lai {\it et al.},
%``Global QCD analysis and the CTEQ parton distributions,''
Phys.\ Rev.\ D {51} (1995) 4763.
%[arXiv:hep-ph/9410404].
%%CITATION = HEP-PH 9410404;%%

%standard model neutrino-nucleon scattering

%\cite{Gandhi:1998ri}
\bibitem{Gandhi:1998ri}
R.~Gandhi, C.~Quigg, M.~H.~Reno and I.~Sarcevic,
%``Neutrino interactions at ultrahigh energies,''
Phys.\ Rev.\ D {58} (1998) 093009.
%[hep-ph/9807264].
%%CITATION = HEP-PH 9807264;%%

%\cite{Gluck:1999js}
\bibitem{Gluck:1999js}
M.~Gl\"uck, S.~Kretzer and E.~Reya,
%``Dynamical QCD predictions for ultrahigh energy neutrino cross sections,''
Astropart.\ Phys.\  {11} (1999) 327;
%[astro-ph/9809273].
%%CITATION = ASTRO-PH 9809273;%%
%
%\cite{Kwiecinski:1999yf}
%\bibitem{Kwiecinski:1999yf}
J.~Kwiecinski, A.~D.~Martin and A.~M.~Stasto,
%``Penetration of the earth by ultrahigh energy neutrinos predicted by low  x QCD,''
Phys.\ Rev.\ D {59} (1999) 093002.
%[astro-ph/9812262].
%%CITATION = ASTRO-PH 9812262;%%

%\cite{Dicus:2001kb}
\bibitem{Dicus:2001kb}
D.~A.~Dicus, S.~Kretzer, W.~W.~Repko and C.~Schmidt,
%``Ultrahigh-energy neutrino nucleon cross-sections and perturbative  unitarity,''
Phys.\ Lett.\ B {514} (2001) 103;
%[hep-ph/0103207].
%%CITATION = HEP-PH 0103207;%%
%
%\cite{Kusenko:2001gj}
%\bibitem{Kusenko:2001gj}
A.~Kusenko and T.~Weiler,
%``Neutrino cross sections at high energies and the future observations of  ultrahigh-energy cosmic rays,''
hep-ph/0106071;
%%CITATION = HEP-PH 0106071;%%
%
%\cite{Reno:2001hv}
%\bibitem{Reno:2001hv}
M.~H.~Reno, I.~Sarcevic, G.~Sterman, M.~Stratmann and W.~Vogelsang,
%``Ultrahigh energy neutrinos, small x and unitarity,''
%arXiv:
hep-ph/0110235.
%%CITATION = HEP-PH 0110235;%%

%Auger sensitivity to horizontal air showers

%\cite{Capelle:1998zz}
\bibitem{Capelle:1998zz}
K.~S.~Capelle, J.~W.~Cronin, G.~Parente and E.~Zas,
%``On the detection of ultra high energy neutrinos with the Auger  Observatory,''
Astropart.\ Phys.\  {8} (1998) 321.
%[astro-ph/9801313].
%%CITATION = ASTRO-PH 9801313;%%


%uhecnu fluxes: reviews

%\cite{Protheroe:1999ei}
\bibitem{Protheroe:1999ei}
R.~J.~Protheroe,
%``High energy neutrino astrophysics,''
Nucl.\ Phys.\ Proc.\ Suppl.\  {77} (1999) 465;
%%CITATION = NUPHZ,77,465;%%
%
%\cite{Gandhi:2000kq}
%\bibitem{Gandhi:2000kq}
R.~Gandhi,
%``Ultra-high energy neutrinos: A review of theoretical and  phenomenological issues,''
Nucl.\ Phys.\ Proc.\ Suppl.\  {91} (2000) 453;
%[hep-ph/0011176].
%%CITATION = HEP-PH 0011176;%%
%
%\cite{Learned:2000sw}
%\bibitem{Learned:2000sw}
J.~G.~Learned and K.~Mannheim,
%``High-energy neutrino astrophysics,''
Ann.\ Rev.\ Nucl.\ Part.\ Sci.\  {50} (2000) 679.
%%CITATION = ARNUA,50,679;%%

%atmospheric neutrinos 

%\cite{Volkova:1980sw}
\bibitem{Volkova:1980sw}
L.~V.~Volkova,
%``Energy Spectra And Angular Distributions Of Atmospheric Neutrinos,''
Sov.\ J.\ Nucl.\ Phys.\  {31} (1980) 784
[Yad.\ Fiz.\  { 31} (1980) 1510];
%%CITATION = SJNCA,31,784;%%
%
%\cite{Lipari:1993hd}
%\bibitem{Lipari:1993hd}
P.~Lipari,
%``Lepton spectra in the earth's atmosphere,''
Astropart.\ Phys.\  {1} (1993) 195.
%%CITATION = APHYE,1,195;%%

%gzk effect

%\cite{Greisen:1966jv}
\bibitem{Greisen:1966jv}
K.~Greisen,
%``End To The Cosmic Ray Spectrum?,''
Phys.\ Rev.\ Lett.\  {16} (1966) 748;
%%CITATION = PRLTA,16,748;%%
%
%\cite{Zatsepin:1966jv}
%\bibitem{Zatsepin:1966jv}
G.~T.~Zatsepin and V.~A.~Kuzmin,
%``Upper Limit Of The Spectrum Of Cosmic Rays,''
JETP Lett.\  {4} (1966) 78
[Pisma Zh.\ Eksp.\ Teor.\ Fiz.\  {4} (1966) 114].
%%CITATION = JTPLA,4,78;%%

%cosmogenic neutrino fluxes

%history

%\cite{Beresinsky:1969qj}
\bibitem{Beresinsky:1969qj}
V.~S.~Berezinsky and G.~T.~Zatsepin,
%``Cosmic Rays At Ultrahigh-Energies (Neutrino?),''
Phys.\ Lett.\ B {28} (1969) 423;
%%CITATION = PHLTA,B28,423;%%
%
%\cite{Beresinsky:1970}
%\bibitem{Beresinsky:1970}
V.~S.~Berezinsky and G.~T.~Zatsepin,
Sov.\ J.\ Nucl.\ Phys.\  {11} (1970) 111
[Yad.\ Fiz.\ {11} (1970) 200].
%%CITATION = SJNCA,11,111;%%

%\cite{Stecker:1979ah}
\bibitem{Stecker:1979ah}
F.~W.~Stecker,
%``Diffuse Fluxes Of Cosmic High-Energy Neutrinos,''
Astrophys.\ J.\  {228} (1979) 919.
%%CITATION = ASJOA,228,919;%%

%\cite{Hill:1985mk}
\bibitem{Hill:1985mk}
C.~T.~Hill and D.~N.~Schramm,
%``The Ultrahigh-Energy Cosmic Ray Spectrum,''
Phys.\ Rev.\ D {31} (1985) 564;
%%CITATION = PHRVA,D31,564;%%
%
%\cite{Hill:1986fm}
%\bibitem{Hill:1986fm}
C.~T.~Hill, D.~N.~Schramm and T.~P.~Walker,
%``Implications Of The Ultrahigh-Energy Cosmic-Ray Spectrum Observed By  The Fly's Eye Detector,''
Phys.\ Rev.\ D {34} (1986) 1622;
%%CITATION = PHRVA,D34,1622;%%
%
%\cite{Stecker:1991vm}
%\bibitem{Stecker:1991vm}
F.~W.~Stecker, C.~Done, M.~H.~Salamon and P.~Sommers,
%``High-energy neutrinos from active galactic nuclei,''
Phys.\ Rev.\ Lett.\  {66} (1991) 2697
[Erratum-ibid.\  { 69} (1991) 2738].
%%CITATION = PRLTA,66,2697;%%

%recent


%\cite{Yoshida:1993pt}
\bibitem{Yoshida:1993pt}
S.~Yoshida and M.~Teshima,
%``Energy spectrum of ultrahigh-energy cosmic rays with extragalactic origin,''
Prog.\ Theor.\ Phys.\  {89} (1993) 833.
%%CITATION = PTPKA,89,833;%%

%\cite{Protheroe:1996ft}
\bibitem{Protheroe:1996ft}
R.~J.~Protheroe and P.~A.~Johnson,
%``Propagation of ultrahigh-energy protons over cosmological distances and implications for topological defect models,''
Astropart.\ Phys.\  {4} (1996) 253
[Erratum-ibid.\  {5} (1996) 215].
%[astro-ph/9506119].
%%CITATION = ASTRO-PH 9506119;%%

%\cite{Yoshida:1997ie}
\bibitem{Yoshida:1997ie}
S.~Yoshida, H.~Dai, C.~C.~Jui and P.~Sommers,
%``Extremely high energy neutrinos and their detection,''
Astrophys.\ J.\  {479} (1997) 547.
%[astro-ph/9608186].
%%CITATION = ASTRO-PH 9608186;%%

%\cite{Engel:2001hd}
\bibitem{Engel:2001hd}
R.~Engel and T.~Stanev,
%``Neutrinos from propagation of ultra-high energy protons,''
Phys.\ Rev.\ D {64} (2001) 093010.
%[arXiv:astro-ph/0101216].
%%CITATION = ASTRO-PH 0101216;%%


%UHECNU flux: upper bound

%\cite{Waxman:1999yy}
\bibitem{Waxman:1999yy}
E.~Waxman and J.~N.~Bahcall,
%``High energy neutrinos from astrophysical sources: An upper bound,''
Phys.\ Rev.\ D {59} (1999) 023002; 
%[arXiv:hep-ph/9807282]. 
%%CITATION = HEP-PH 9807282;%%
%
%\cite{Bahcall:2001yr}
%\bibitem{Bahcall:2001yr}
J.~N.~Bahcall and E.~Waxman,
%``High energy astrophysical neutrinos: The upper bound is robust,''
Phys.\ Rev.\ D {64} (2001) 023002.
%[arXiv:hep-ph/9902383].
%%CITATION = HEP-PH 9902383;%%

%\cite{Mannheim:2001wp}
\bibitem{Mannheim:2001wp}
K.~Mannheim, R.~J.~Protheroe and J.~P.~Rachen,
%``On the cosmic ray bound for models of extragalactic neutrino  production,''
Phys.\ Rev.\ D {63} (2001) 023003.
%[arXiv:astro-ph/9812398].
%%CITATION = ASTRO-PH 9812398;%%


%\cite{Berezinsky:1979pd}
\bibitem{Berezinsky:1979pd}
V.~S.~Berezinsky,
%``High-Energy Neutrino Astronomy Versus Gamma Astronomy. (Talk),''
in: J.~G.~Learned (Ed.), DUMAND Summer Workshops,  
Khabarovsk and Lake Baikal, 22-31 Aug 1979, Hawaii DUMAND Center, 
University of Hawaii, 1980, pp. 245-261.

%Z-burst scenario

%\cite{Fargion:1999ft}
\bibitem{Fargion:1999ft}
D.~Fargion, B.~Mele and A.~Salis,
%``Ultrahigh energy neutrino scattering onto relic light neutrinos in  galactic 
%halo as a possible source of highest energy extragalactic  cosmic rays,''
Astrophys.\ J.\  {517} (1999) 725; 
%[arXiv:astro-ph/9710029].
%%CITATION = ASTRO-PH 9710029;%%
%
%\cite{Weiler:1999sh}
%\bibitem{Weiler:1999sh}
T.~J.~Weiler,
%``Cosmic ray neutrino annihilation on relic neutrinos revisited:  
%A mechanism for generating air showers above the  Greisen-Zatsepin-Kuzmin cut-off,''
Astropart.\ Phys.\  {11} (1999) 303; 
%[arXiv:hep-ph/9710431].
%%CITATION = HEP-PH 9710431;%%
%
%\cite{Yoshida:1998it}
%\bibitem{Yoshida:1998it}
S.~Yoshida, G.~Sigl and S.~J.~Lee,
%``Extremely high energy neutrinos, neutrino hot dark matter, and the  highest energy cosmic rays,''
Phys.\ Rev.\ Lett.\  {81} (1998) 5505; 
%[arXiv:hep-ph/9808324].
%%CITATION = HEP-PH 9808324;%%
%
%\cite{Fodor:2001qy}
%\bibitem{Fodor:2001qy}
Z.~Fodor, S.~D.~Katz and A.~Ringwald,
%``Determination of absolute neutrino masses from Z-bursts,''
%arXiv:
hep-ph/0105064.
%%CITATION = HEP-PH 0105064;%%


%Interpretation of Fly's Eye bound

%\cite{Reno:1988zf}
\bibitem{Reno:1988zf}
M.~H.~Reno and C.~Quigg,
%``On The Detection Of Ultrahigh-Energy Neutrinos,''
Phys.\ Rev.\ D {37} (1988) 657.
%%CITATION = PHRVA,D37,657;%%

%history: bound on neutrino-nucleon cross section

%\cite{Berezinsky:1974kz}
\bibitem{Berezinsky:1974kz}
V.~S.~Berezinsky and A.~Y.~Smirnov,
%``Astrophysical Upper Bounds On Neutrino-Nucleon Cross Section At Energy E>=3 X 10-To-The-17 Ev,''
Phys.\ Lett.\ B {48} (1974) 269.
%%CITATION = PHLTA,B48,269;%%


%RICE

%\cite{Frichter:1999kr}
\bibitem{Frichter:1999kr}
G.~Frichter  [RICE Collaboration. AMANDA Collaboration],
%``Status of the RICE experiment,''
prepared for: B.L. Dingus, D.B. Kieda, M.H. Salamon (Eds.),  
26th International Cosmic Ray Conference (ICRC 99), Salt Lake City, UT, 17-25 Aug 1999,
Melville, AIP, 2000.

\end{thebibliography}
\end{document}